# Circular photogalvanic effect in 2D van der Waals heterostructure

Abdullah Rasmita[1†], Chongyun Jiang[1, 2†*], Hui Ma[1], Zhurun Ji[3], Ritesh Agarwal[3] & Wei-bo Gao[1,4*]

[1]Division of Physics and Applied Physics, School of Physical and Mathematical Sciences, Nanyang Technological University, Singapore 637371, Singapore
[2]College of Electronic Information and Optical Engineering, Nankai University, Tianjin, China
[3]Department of Materials Science and Engineering, University of Pennsylvania, Philadelphia, PA 19104, USA
[4]The Photonics Institute and Centre for Disruptive Photonic Technologies, Nanyang Technological University, 637371 Singapore, Singapore

*Corresponding authors: Chongyun Jiang, Wei-bo Gao

†Authors contributed equally

**Abstract**

Utilizing spin or valley degree of freedom is one of the promising approaches to realize more energy-efficient information processing. In the 2D transition metal dichalcogenide, the spin/valley current can be generated by utilizing the circular photogalvanic effect (CPGE), i.e., the generation of photocurrent by a circularly polarized light. Here we show that an in-plane electric field at $MoS_2/WSe_2$ heterostructure-electrode boundary results in an electrically tunable circular photogalvanic effect (CPGE) under optical excitation with normal incidence. The observed CPGE can be explained by the valence band shift due to the in-plane electric field and different effective relaxation times between hole and electron combined with the valley optical selection rule. Furthermore, we show that the CPGE can be controlled by changing the Fermi level using an out-of-plane electric field. Such phenomena persists even at room temperature. This finding may facilitate the utilization of 2D heterostructure as an opto-valleytronics and opto-spintronics device platform.

**Introduction**

Manipulation of the spin degree of freedom or, equivalently, pseudo-spin such as valley degree of freedom can result in practical information processing devices that are more energy-efficient compared to the currently utilized charge-based approach (1-3). The two-dimensional van der Waals heterostructure has emerged as a potential material class for realizing the future electronics and spintronics/valleytronics devices (4-8). This is due to the possibility to engineer the interlayer interaction (9-14) and to tailor the properties of different materials through proximity effect (5, 15-20). In this regard, it is important to study how the spin/valley current can be generated in 2D van der Waals heterostructure.

Due to the large spin-orbit coupling and the valley optical selection rule in monolayer transition metal dichalcogenide (TMD) (21-26), the spin and valley current can be generated optically by utilizing phenomena such as the valley Hall effect (27-29) and the circular photogalvanic effect (CPGE) (30-32). CPGE is the generation of photocurrent by circularly polarized light. In CPGE, two different circular polarization results in charge currents with a different magnitude or/and direction. In contrast with the valley Hall effect, a nonzero longitudinal charge current is not needed in the CPGE. This results in less dissipation compared to the valley Hall-based method (27, 28, 33).

Additionally, CPGE can be regarded as a tool to probe the material symmetry. This type of tool is particularly important for studying topological properties since topology in solid has a close relationship with symmetry (34, 35). The existence of CPGE requires that the inversion symmetry is broken (30, 36, 37). In particular, the observation of CPGE under normal incidence requires a lower order in-plane symmetry where only a single mirror symmetry is allowed (30, 36). This phenomenon has been observed in some 2D Weyl semimetals such as $WTe_2$ (32),

MoTe$_2$, and Mo$_{0.9}$W$_{0.1}$Te$_2$ (38). In the monolayer WTe$_2$ case, the low order symmetry is intrinsically inherent and the CPGE is closely related to the topological properties in this material through the Berry curvature dipole (39) while in the MoTe$_2$, and Mo$_{0.9}$W$_{0.1}$Te$_2$ case it is related to the asymmetric carrier generation due to Gaussian profile optical excitation (38).

Here we report a new mechanism for which the normal incidence excitation can generate the CPGE. We show that the CPGE can be observed at MoS$_2$/WSe$_2$ heterostructure-electrode boundary where there is a built-in in-plane electric field due to the laterally extending Schottky depletion region (40, 41). The CPGE polarity depends on the in-plane electric field direction, which can be changed by an external modulation. Indeed, we show that the source-drain bias can modulate both the magnitude and the polarity of the CPGE current. Moreover, we show that the back gating can control the magnitude of the CPGE current. We also propose a theoretical explanation for the generated CPGE. Specifically, we show that the in-plane electric field shifts the valence band of WSe$_2$. In particular, the K and K' valley shift in the opposite direction. As the electron and hole have different effective relaxation time in the heterostructure, this will result in a non-zero valley-dependent photocurrent which has opposite direction in K and K' valleys. Combined with the valley optical selection rules, CPGE current can then be generated. Our simulation shows a good agreement with the experimentally obtained CPGE pattern. To the best of our knowledge, our finding is the first demonstration of CPGE current induced by an in-plane electric field in 2D TMD heterostructure.

**Results and Discussion**

Fig. 1*A* shows a microscopic image of the fabricated device. It consists of gold electrodes patterned on top of MoS$_2$/WSe$_2$ heterostructure with SiO$_2$/Si as the substrate. The structure of the MoS$_2$/WSe$_2$ sample is illustrated in Fig. 1*B*. Both monolayer MoS$_2$ and monolayer WSe$_2$ consist of Mo(W) atom located between the S(Se) atomic layer, creating hexagonal lattice and broken inversion symmetry. Due to the broken inversion symmetry, the Berry curvature becomes nonzero, which results in the light with different circular polarization couples to different valleys (21-26). As illustrated in Fig. 1*C*, the type-II band alignment of MoS$_2$/WSe$_2$ causes electrons to relax to the MoS$_2$ layer while holes relax to the WSe$_2$ layer. This mechanism suppresses the electron-hole exchange interaction. Since the exciton intervalley scattering depends on the electron-hole exchange interaction (42-44), this results in the suppression of the intervalley scattering even at room temperature (45, 46). Moreover, comparing the heterostructure with a monolayer, the suppression of exciton intervalley scattering could increase the carrier relaxation time, and thus the CPGE current.

The setup for detecting and characterizing the CPGE is shown in Fig. 1*D*. The polarization state of the light is controlled either by changing the phase retardation of the liquid crystal modulator or by changing the quarter-wave plate (QWP) axis orientation. The excitation can be scanned over the sample area by stirring the beam. In the case that liquid crystal is used to modulate the circular polarization, the current is passed into a lock-in amplifier to get the CPGE current, which is defined as the difference between the source-drain current under left-circularly ($\sigma_+$) and right-circularly ($\sigma_-$) polarized excitation: $I_{CPGE} = I_{SD}(\sigma_+) - I_{SD}(\sigma_-)$. The source-drain voltage, $V_{SD}$, and the back gate voltage, $V_G$, can be varied independently. All of the experiments were done under normal incidence, which rules out any contribution from the circular photon drag effect (30).

First, we obtained the CPGE intensity map (i.e. $I_{CPGE}$, vs. laser spot location) at room temperature (295 K) under zero bias ($V_{SD} = V_G = 0\text{V}$) by modulating the circular polarization using liquid crystal modulator and by scanning the beam throughout the sample. The results for two different electrode configurations are shown in Fig. 1*E* and 1*F*, respectively. The reproducibility of the phenomenon is confirmed by testing a different sample using the same setup. The CPGE is much more apparent for excitation near the edge of the metal electrodes compared to the other parts of the sample. This observation is correct regardless of the electrode pair configuration and the temperature (see SI Section I and SI Fig. S1 and S2). This shows that the CPGE depends on parameters that are large only near the electrode edges, such as a built-in electric field.

To verify if the electric field plays a significant role in the observed CPGE, we study the effect of the source-drain voltage on the CPGE. As shown in the COMSOL simulation (see SI Fig. S3), the source-drain voltage can affect both the in-plane and the out-of-plane electric field. However, since the $MoS_2$/$WSe_2$ is more conducting than $SiO_2$, the source-drain voltage will mainly change the in-plane electric field. Comparing Fig. 2A(B) and 1E(F), we can see that the CPGE is higher at a lower temperature. Hence, to get a higher signal-to-noise ratio, we study the source-drain voltage dependence under low temperature (140 K) instead of room temperature. In Fig. 2C, the photocurrent at various source-drain voltage under the temperature of 140K is plotted as a function of quarter-wave plate fast axis angle with respect to the polarizer axis. The dark current is subtracted by using a chopper and a lock-in amplifier. The data is fitted using the fitting function

$$I_{SD} = I_0 + I_1 \sin(4\theta) + I_2 \cos(4\theta) + I_{CPGE} \sin(2\theta) .$$  [1]

Here, $I_0$ is the polarization-independent component while $I_1$ and $I_2$ are related to the linear photogalvanic effect. In Fig. 2D, the $I_{CPGE}$ obtained from the fitting is plotted as a function of source-drain voltage. From this figure, we can see that the source-drain voltage can change both the magnitude and the polarity of the CPGE current.

Next, we study the back gate voltage dependence of the photocurrent, which mainly affects the out-of-plane electric field. For this study, we found that the signal-to-noise ratio is good enough to observe the relationship at room temperature. Initially, we use the liquid crystal modulator to modulate the excitation polarization between $\sigma_+$ and $\sigma_-$ polarization. The plot of $I_{CPGE}$ as a function of the back gate voltage $V_G$ taken at room temperature (295 K) is shown in Fig. 3A upper panel. As can be seen from this figure, the $I_{CPGE}$ magnitude can be changed by modifying the back gate voltage. Additionally, we also study the polarization-independent part (including the dark current) which is defined as $\overline{I_{SD}} = \frac{I_{SD}(\sigma_+) + I_{SD}(\sigma_-)}{2}$. The plot of $\overline{I_{SD}}$ as a function of source-drain voltage is shown in Fig. 3A lower panel. Comparing Fig. 3A upper and lower panel, we can see that the back gate voltage modulates the magnitude of both $I_{CPGE}$ and $\overline{I_{SD}}$ in the same way.

We further confirm this by replacing the liquid crystal modulator with a quarter-wave plate to study how the back gate voltage affects the excitation polarization dependence of the source-drain current. The result is shown in Fig. 3B. Using the fitting function described in Eq. **1**, we can obtain the polarization-independent, linear, and circular photogalvanic current. These three types of photocurrent can then be plotted as a function of the back gate voltage, as shown in Fig. 3C. As can be seen from this figure, the back gate voltage modulates the magnitude of the photocurrent. Furthermore, regardless of the photocurrent type (polarization-independent, linear, and circular), all of them show the same trend of going to zero as the back gate voltage becomes more negative. Further characterization indicates that the CPGE magnitude has a linear dependence on excitation power and maximum when the excitation wavelength is near-resonant with the $WSe_2$ exciton transition (see SI Section II and SI Fig. S4).

The experiments above show that the source-drain voltage affects both the magnitude and polarity of the CPGE while the back gate voltage modulates the CPGE magnitude. Hence, we can conclude that the in-plane electric field affects both the CPGE magnitude and polarity while the out-of-plane electric field is mainly affecting the CPGE magnitude. Additionally, we found that the back gate voltage affects all types of photocurrent in the same way regardless of the excitation polarization. In the next part, we discuss the theoretical explanation of these experimental results.

We start by analyzing the effect of the in-plane electric field on the photocurrent. A detailed discussion can be found in SI Section IV-VI. Here, we give a simple description of the proposed mechanism which is illustrated in Fig. 4a for the case of near-resonant excitation (i.e., the excitation near $\vec{k}=0$). We first note that the valley optical selection rule in WSe$_2$ leads to the result that the light with $\sigma_+$ polarization can only generate carrier in K valley while that with $\sigma_-$ polarization can only generate carrier in K' valley. Hence, a nonzero CPGE current corresponds to a nonzero difference between the valley-dependent photocurrent in K and K' valley. We next explain how the in-plane electric field can generate this difference.

As shown in Fig. 4a, the nonzero in-plane electric field causes the valence band at K and K' valley to shift in two opposite directions. The optical excitation creates non-equilibrium electron and hole with velocity given by $\vec{v}_{e(h)} = \frac{1}{\hbar} \nabla_k E_{VB(CB)}(\vec{k})$ where $\vec{v}_{e(h)}$ is the electron (hole) velocity and the $E_{VB(CB)}(\vec{k})$ is the $\vec{k}$-dependent valence (conduction) band energy. This means that the electron and hole velocities are proportional to the band slope. Referring to Fig. 4a and considering only one valley, for near-resonant excitation, there are two transitions with the same energy. Each of these two transitions generates either nonzero electron velocity or hole velocity. In MoS$_2$/WSe$_2$ heterostructure, the electron will undergo ultrafast charge transfer to the MoS$_2$ layer. Since both the electron relaxation time and the conduction band curvature of MoS$_2$ are different than those in WSe$_2$, the current generated by electron and holes will not cancel each other and a nonzero photocurrent is generated in K/K' valley. Due to the opposite shift between the valence band in the K and K' valley, the photocurrent will have an opposite direction for these two valleys, which results in a nonzero CPGE current. Furthermore, the shift is linearly proportional to the in-plane electric field. Hence, the CPGE current will also be linearly proportional to this field.

More precisely, for CPGE current calculation, the system can be modeled as a two-band system with different effective relaxation time between electron and hole (see SI Section IV-VI). Following this model, the CPGE current density, $\vec{J}_{CPGE}$, can be expressed as

$$\vec{J}_{CPGE} = A(\hat{x} C_1 \sin\alpha + \hat{y} C_2 \cos\alpha) \quad [2]$$

where $A$ is the electric field amplitude at the excitation location, $C_{1(2)}$ is a constant of the electric field, and $\alpha$ is the angle between the in-plane electric field direction and the zigzag direction (x-axis, see SI Fig. S6). In this expression, the electric field can be written as $\vec{E} = A(\hat{x}\cos\alpha + \hat{y}\sin\alpha)$. The nonzero $\vec{J}_{CPGE}$ shows that the electric field can break the overall system symmetry.

Now we discuss how the CPGE map pattern (Fig. 1(E, F), Fig. 2(A, B), and SI Fig. S1 and S2) can be explained theoretically. First, the built-in electric field magnitude is only significant near the electrode edge, so the $\vec{J}_{CPGE}$ (and thus $I_{CPGE}$) is negligible for excitation far away from the edge. Next, regarding the pattern near the electrode edge, the formulation in (47) can be used to obtain $I_{CPGE}$ from $\vec{J}_{CPGE}$ as

$$I_{CPGE} = A\left(\frac{C_1 + C_2}{2}\right)\sin(2\alpha) \quad [3]$$

For lattice symmetry lower than $C_{3v}$ near the electrode edge, $C_1 + C_2$ is nonzero (see SI Section IV and VI). The simulation of the CPGE current around the electrode is shown in Fig. 4b. In particular, at the two sides of the same electrode, $\alpha$ differs by 180⁰, which results in the same sign for the CPGE current, while between the two perpendicular sides of the electrode $\alpha$ differs by $90^0$, which results in the opposite sign for the CPGE current. In summary, the CPGE map pattern can be well explained by the equation **[3]**.

We now discuss the effect of back-gate voltage (the out-of-plane electric field). Applying a back-gate voltage is equivalent to carrier doping, which will change the Fermi level. There are two effects of changing the Fermi level. The first one is that it will change the built-in electric field since it affects the band bending between the MoS$_2$(WSe$_2$) and the Cr/Au electrode. This will affect the $I_{CPGE}$ in the same way as the in-plane electric field does. The second effect is that it can change the conduction band electron mobility. When the Fermi level is higher than the energy of some defect states, the defect will be fully occupied and cannot act as a scattering center (48). This results in higher conduction band electron mobility. Unlike the first mechanism, the second one will affect all types of current. The interplay between these two mechanisms depends on the energy level of the traps. However, based on the fact that the $I_{CPGE}$ has the same tendency as the $\overline{I_{SD}}$, we conclude that the second mechanism is more dominant in our case.

In summary, we have experimentally demonstrated CPGE under excitation with a normal incidence near MoS$_2$/WSe$_2$ heterostructure-electrode interface. The CPGE can be controlled by the source-drain bias and the electrical back gating. The source-drain dependence can be explained by the valence band shift due to the in-plane electric field, valley-specific optical selection rules, and the asymmetry in electron and hole effective relaxation time. The theoretical simulation fits very well with the experimental results for the CPGE map patterns. The back gate dependence is attributed to the modulation of the doping level, which affects the charge trap occupation and, hence, the carrier relaxation time. The control by source-drain bias opens up new possibilities for new devices based on planar electrode design. Combined with the long carrier and spin polarization lifetime in 2D heterostructure (49, 50), this finding may be utilized to realize a practical valleytronics and spintronics semiconductor devices.

**Materials and Methods**

The MoS$_2$/WSe$_2$ heterostructure is fabricated via the all-dry transfer method. MoS$_2$ and WSe$_2$ monolayer flakes are primarily exfoliated on polydimethylsiloxane (PDMS) stamps. Then the PDMS stamps are sequentially aligned and transferred onto a SiO$_2$/Si substrate. After vacuum annealing at 150℃, the ohmic electrodes are fabricated concurrently connecting both layers using standard electron beam lithography (EBL) and thermal evaporation techniques, which consist of 5 nm Cr and 50 nm Au.

**Acknowledgments**

We thank Wang Yao, Feng Wang, and J.C.W. Song for the discussion. We acknowledge the financial support from the Singapore National Research Foundation through a Singapore 2015 NRF fellowship grant (NRF-NRFF2015-03), Singapore Ministry of Education (MOE2016-T2-2-077 and MOE2011-T3-1-005), A*Star QTE programme.

**Figures and Tables**

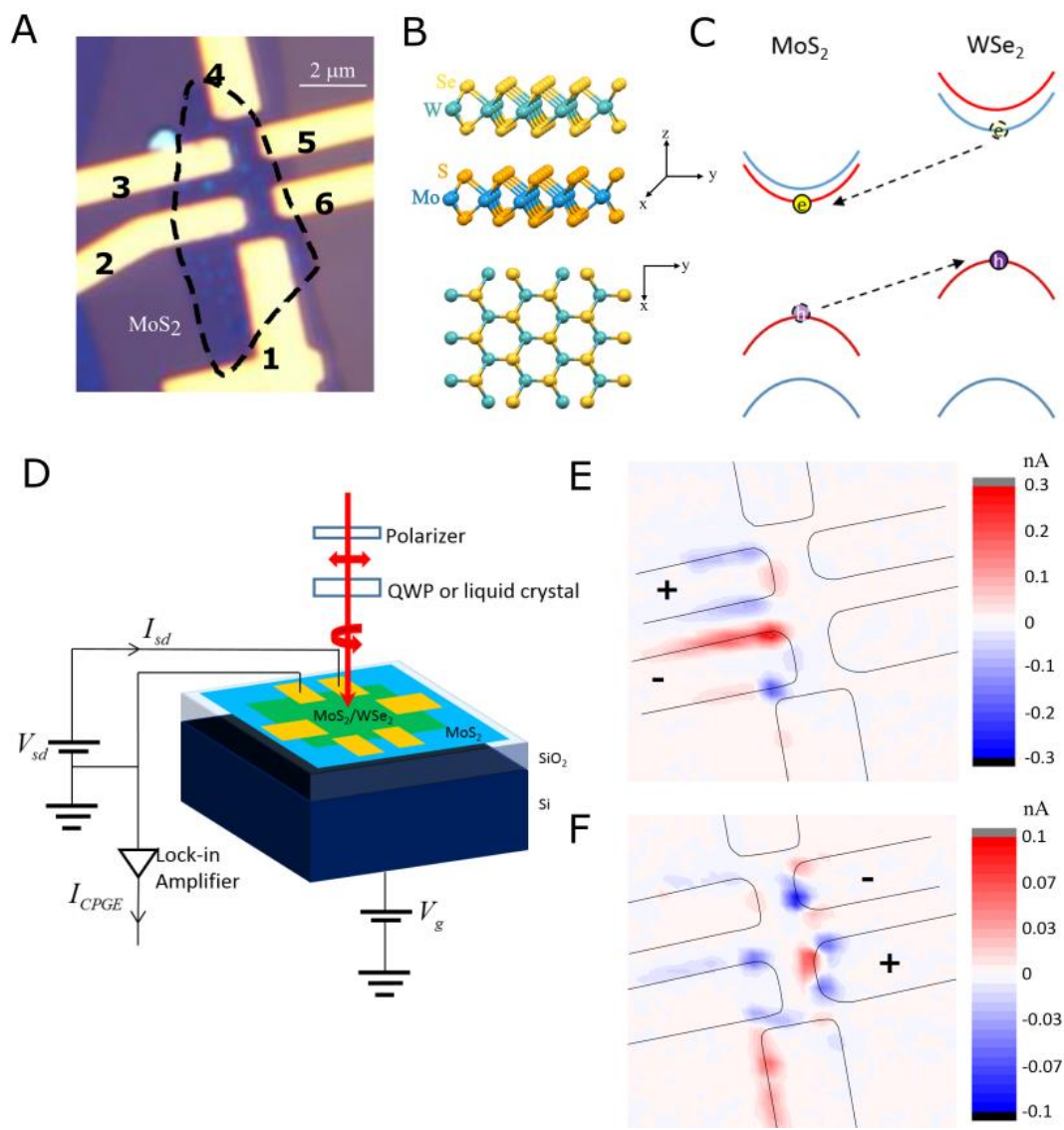

**Figure 1.** Device, experimental setup, and room temperature CPGE intensity map. (*A*) Optical image of the device. The device consists of gold electrodes patterned on to of WSe$_2$/MoS$_2$ heterostructure with SiO$_2$/ Si as the substrate. The number shown is for easy reference to the electrodes. (*B*) Heterostructure crystal structure. It consists of monolayer WSe$_2$ stacked on top of monolayer MoS$_2$. (*C*) The carrier relaxation in the heterostructure. The electrons relax quickly to the MoS$_2$ layer while the holes relax to the WSe$_2$ layer reducing the electron-hole exchange interaction. (*D*) Experimental setup. (*E*) CPGE map with source: electrode 3 and drain: electrode 2. The CPGE current is defined as the difference between the source-drain current under $\sigma_+$ and $\sigma_-$ excitation: $I_{CPGE} = I_{SD}(\sigma_+) - I_{SD}(\sigma_-)$. (*F*) CPGE map with source: electrode 6 and drain: electrode 5. Both CPGE map is obtained using 720 nm 115 μW excitation at room temperature (295 K).

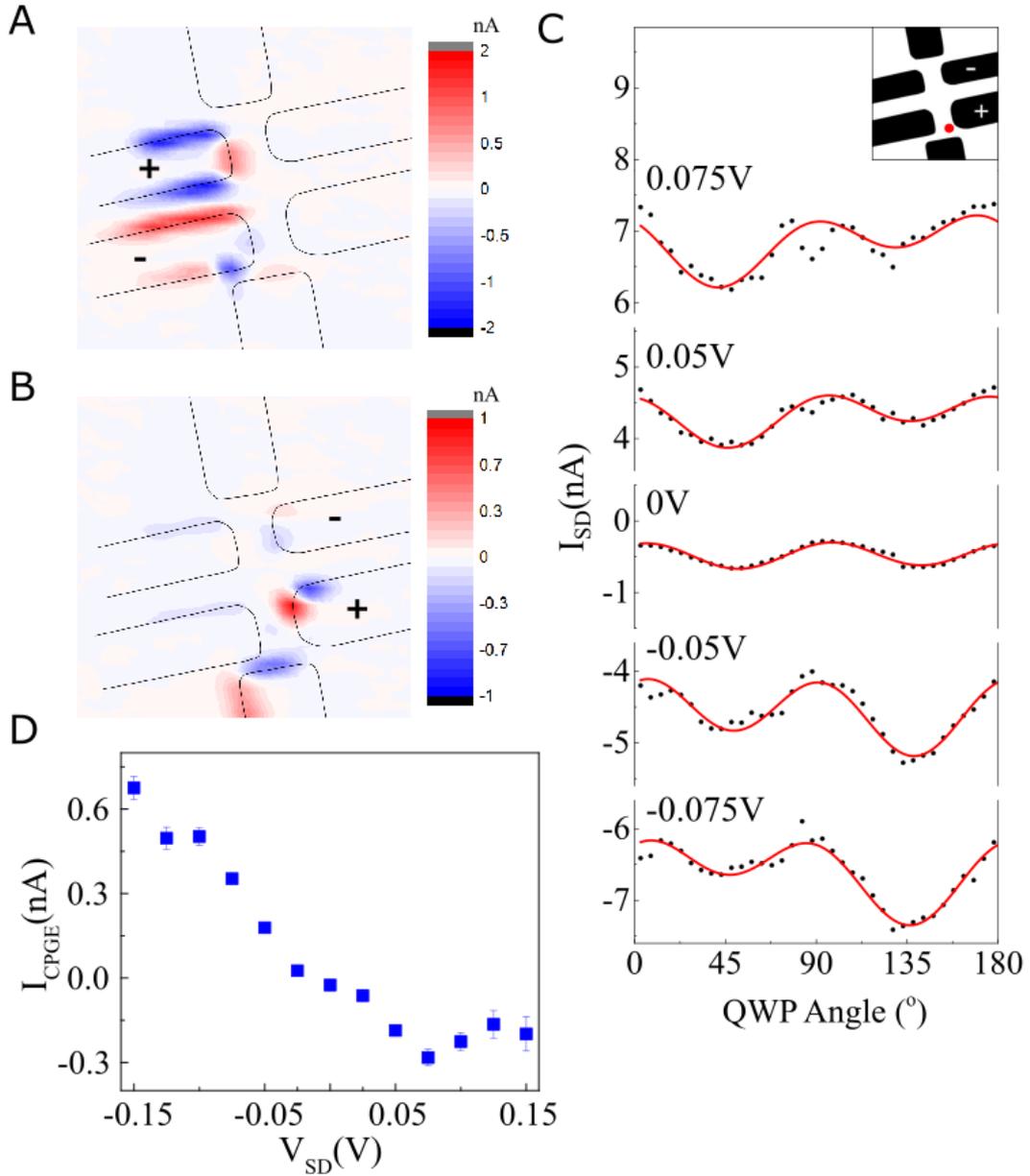

**Figure 2.** CPGE map and characterization against source-drain voltage at 140 K. (*A*) CPGE map with source: electrode 3 and drain: electrode 2. (*B*) CPGE map with source: electrode 6 and drain: electrode 5. (*C*) Excitation polarization effect on the source-drain current at various source-drain voltage. The source-drain current is plotted as a function of quarter wave plate fast axis angle with respect to the polarizer axis. Only the photocurrent is measured. The dark current is subtracted by using a chopper and a lock-in amplifier. (*D*) CPGE as a function of source-drain voltage. The results are obtained by fitting the data in Fig. 2*C* using Eq. **1**. All of the data is collected using 720 nm 100 μW excitation.

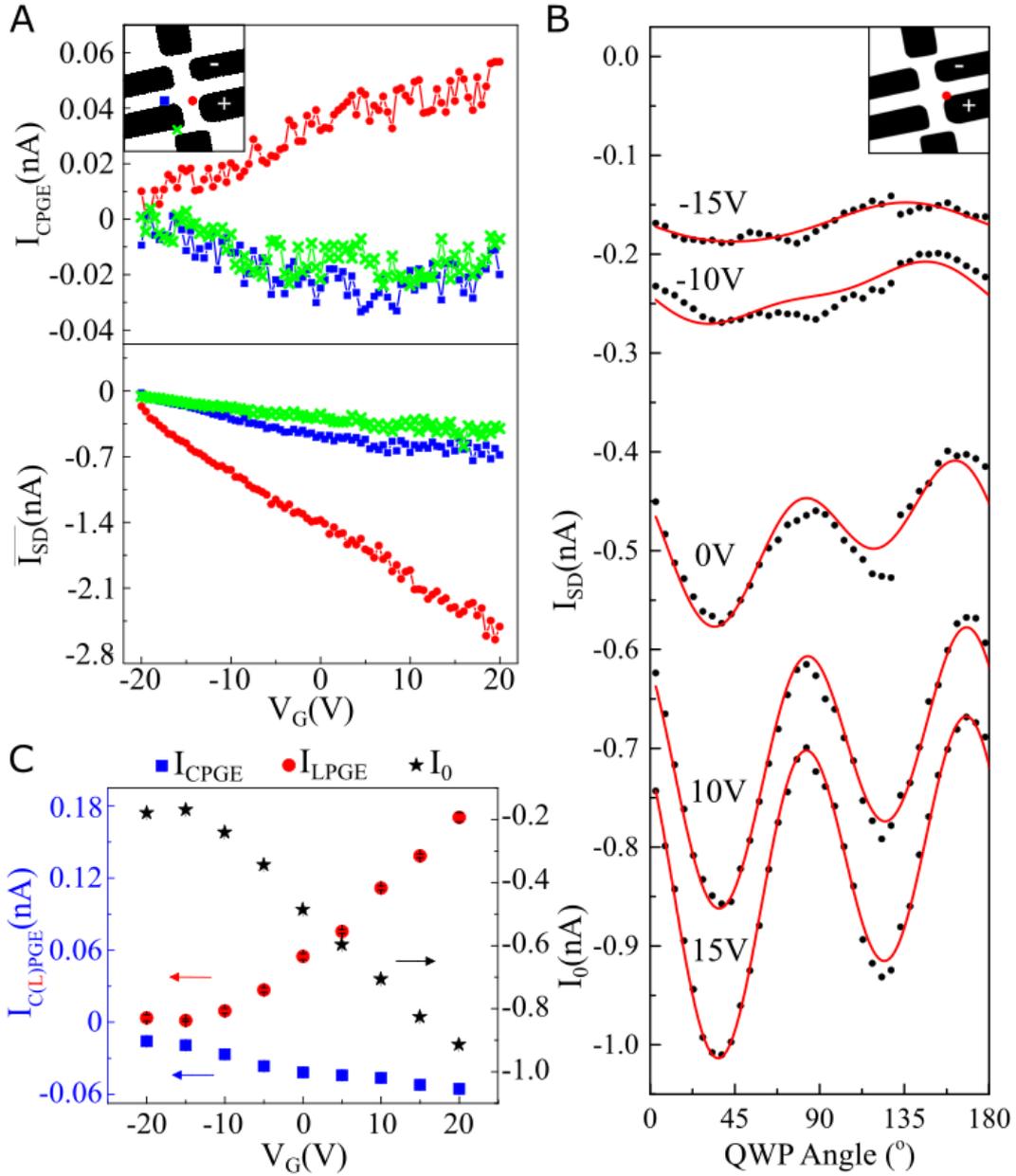

**Figure 3.** CPGE characterization against gate voltage at 295 K. (*A*) CPGE current as a function of gate voltage. A liquid crystal modulator is used to modulate between $\sigma_+$ and $\sigma_-$ excitation. The $\overline{I_{SD}}$ is defined as $\overline{I_{SD}} = \dfrac{I_{SD}(\sigma_+) + I_{SD}(\sigma_-)}{2}$. The data is obtained using a 720 nm 115 μW laser. (*B*) Excitation polarization effect on the source-drain current at various gate voltage. The source-drain current is plotted as a function of quarter wave plate fast axis angle with respect to the polarizer axis. Only the photocurrent is measured. The dark current is subtracted by using a chopper and a lock-in amplifier. (*C*) CPGE, LPGE, and polarization-independent current as a function of the source-drain voltage. The results are obtained by fitting the data in Fig. 3*B* using Eq. **1** with $I_{LPGE} = \sqrt{I_1^2 + I_2^2} \approx I_1$. The data is collected using a 720 nm 115 μW laser excitation.

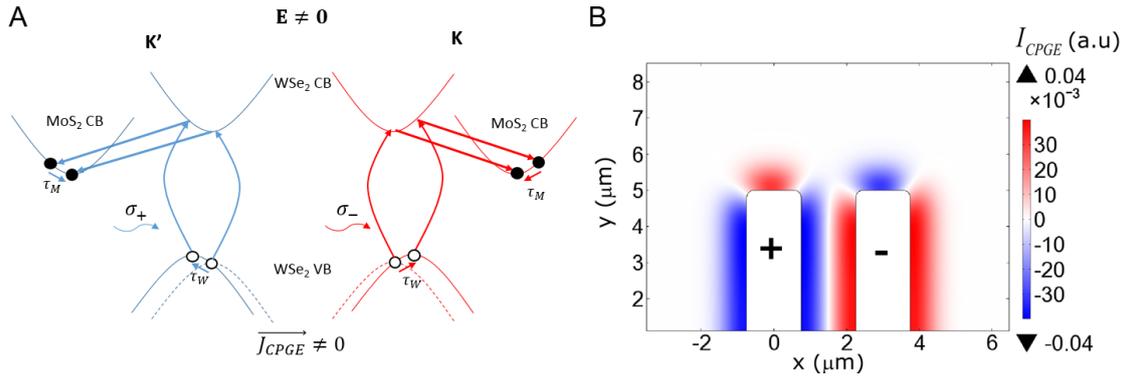

**Figure 4.** Microscopic model of the CPGE generation and the simulation result. (*A*) The in-plane electric field shifts the valence band at K and K' valley in two opposite directions. Resonant optical excitation creates electron and hole with the same nonzero velocity. Due to the difference in carrier relaxation time in $MoS_2$ and $WSe_2$, the hole and electron will have different effective relaxation time. Hence, a nonzero valley-dependent photocurrent is generated. The photocurrent has an opposite direction between K and K' valley. Combined with the valley optical selection rule, this results in electrically tunable CPGE current density. (*B*) COMSOL simulation of the $I_{CPGE}$ following equation **[3]**. Here the built-in electric field decays exponentially from the edge with a decay constant of 100 nm and the beam is taken to be Gaussian beam with beam full width at half maximum equal to 1 μm. The angle between the positive x-axis direction and the zigzag direction of the $WSe_2$ is $30^0$.

# Supporting Information for:
# Circular photogalvanic effect in 2D heterostructure

**I. CPGE map for various electrode pair configuration**

The CPGE maps at room temperature for various electrode pair configurations are shown in Fig. S1(*A-I*), while the maps at low temperature (140K) are shown in Fig. S2(*A-I*). As can be seen from these figures, the observation that the $I_{CPGE}$ is maximum near the edge of the electrode is valid regardless of the electrode pair configuration. Also, the CPGE magnitude at 140 K is much larger than that at 295 K. This can be attributed to higher carrier mobility (longer relaxation time) at a lower temperature.

**II. Power and excitation wavelength dependence of CPGE**

The $I_{CPGE}$ grows linearly with the population of the photo-induced carrier, which in turn depends on the excitation power and wavelength. The power and wavelength dependence of the $I_{CPGE}$ at temperature 140 K is shown in Fig. S4*A* and S4*B*, respectively. As expected, the $I_{CPGE}$ grows linearly with increasing excitation power.

It reaches a maximum magnitude at excitation wavelength ~720 nm (i.e., the charged WSe$_2$ exciton peak (1)). The sharp decrease of CPGE magnitude for excitation above 720 nm is due to the fact that excitation with these wavelengths does not have enough energy to excite the intralayer exciton. As a result, there is no photo-induced carrier generated. The decrease in CPGE magnitude for excitation below 720 nm can be understood by considering that off-resonant excitation results in a larger intervalley scattering of the intralayer exciton (2). This results in weaker correspondence between circular polarization and carrier valley polarization. Hence, a reduction in CPGE magnitude is observed.

**III. CPGE map on another MoS$_2$/WSe$_2$ heterostructure device**

To check the reproducibility of the result, we fabricated another MoS$_2$/WSe$_2$ device and took the CPGE map under zero source-drain bias. For this sample, we found that the CPGE map is more apparent when we apply 5V gate voltage. The optical image of the sample and the CPGE map between one electrode pair configuration is shown in Fig. S5(*A, B*). From this figure, we can see that, qualitatively, the conclusion that we obtained from the other device is still applicable here. In particular, it can be seen that the two orthogonal sides of the same electrode have opposite CPGE polarity while the parallel sides have the same CPGE polarity.

**IV. Energy band and Berry curvature modification by a uniform in-plane electric field**

In this section, we derive the energy band and Berry curvature as a function of wave vector $\vec{k}$ for a monolayer TMD under a uniform in-plane electric field. This section is divided into two parts. In the first part, we derive the $\vec{k}$-dependent perturbation Hamiltonian. In the second part, we give the expression of the dispersion relation and Berry curvature for the general case of linear perturbation Hamiltonian and apply it to the case of a small uniform in-plane electric field. In particular, we show that the small linear perturbation cannot generate nonzero Berry curvature dipole.

Throughout this section and the sections afterward, the wave vector $\vec{k}$ is expressed with respect to K(K') valley. Moreover, we only consider the terms linear in $\vec{k}$ in the Hamiltonian. This is acceptable for near-resonant excitation since the $k \approx 0$ in this case.

A. Perturbation Hamiltonian

Near the K(K') valley, the electronic Bloch state of monolayer TMD mainly consists of the $d_{z^2}$ and $\frac{1}{\sqrt{2}}\left(d_{x^2-y^2}+i\tau d_{xy}\right)$ orbital with $\tau$ is the valley index and has the value of $\tau = 1(-1)$ for K(K') valley (3, 4). The Bloch state corresponding to these two orbitals can be written as

$$\left|\Phi_1(\vec{k},\vec{r})\right\rangle = \frac{1}{\sqrt{N}}\sum_{\vec{R}} e^{i\vec{k}\cdot\vec{R}} \left|d_{z^2}(\vec{r}-\vec{R})\right\rangle \text{ and}$$

$$\left|\Phi_2^\tau(\vec{k},\vec{r})\right\rangle = \frac{1}{\sqrt{2N}}\sum_{\vec{R}} e^{i\vec{k}\cdot\vec{R}} \left\{\left|d_{x^2-y^2}(\vec{r}-\vec{R})\right\rangle + i\tau\left|d_{xy}(\vec{r}-\vec{R})\right\rangle\right\}$$

with the summation index $\vec{R}$ is over all possible lattice vectors and $N$ is the number of the lattice sites.

Using $\left\{\left|\Phi_1(\vec{k},\vec{r})\right\rangle, \left|\Phi_2^\tau(\vec{k},\vec{r})\right\rangle\right\}$ as a basis with can get the $\vec{k}$-dependent Hamiltonian. Here, we assume the orthogonal tight-binding assumption: $\left\langle d_i(\vec{r})\middle|d_j(\vec{r}-\vec{R})\right\rangle = \delta_{ij}\delta(\vec{R})$. When the in-plane electric field is zero, up to first order in $\vec{k}$, the Hamiltonian can be expressed as (3, 4)

$$H_0(\vec{k}) = \begin{bmatrix} \Delta/2 & a_0 t(\tau k_x - ik_y) \\ a_0 t(\tau k_x + ik_y) & -\Delta/2 \end{bmatrix} \quad \text{[S4.1]}$$

with $a_0$ is lattice constant and $t$ is the effective hopping integral.

The effect of the in-plane electric field can be obtained by projecting the electric potential onto space with this basis also. We note that, up to a phase, the basis is periodic in space while the electric potential is not. The electric potential can then be separated into two parts: the periodic part and the non-periodic part with the periodic part satisfies $V(\vec{r}+\vec{R}) = V(\vec{r})$. The non-periodic part results in the mixing between Bloch states (i.e., net charge current). The periodic part of the potential will affect the band structure.

Considering the periodic part only, the electric potential within one unit cell due to the in-plane electric field can be expressed as $V(\vec{r}) = V(x,y) = eA(x\cos\alpha + y\sin\alpha)$. Here, $e$ is the electron charge, $(A,\alpha)$ represents electric field amplitude and direction, and $(x, y)$ represents a position in space with respect to the unit cell center (see Fig. S6). Projecting this onto the $\left\{\left|\Phi_1(\vec{k},\vec{r})\right\rangle, \left|\Phi_2^\tau(\vec{k},\vec{r})\right\rangle\right\}$ basis results in the perturbation Hamiltonian $V(\vec{k})$ with the element $V_{ij}(\vec{k}) = \left\langle\Phi_i(\vec{k},\vec{r})\middle|V(\vec{r})\middle|\Phi_j^\tau(\vec{k},\vec{r})\right\rangle$. Within nearest neighbor approximation and keeping terms up to first order in $\vec{k}$, we obtain

$$V_{11}(\vec{k}) = 0 \quad \text{[S4.2]}$$

$$V_{22}(\vec{k}) = 2\tau eAa_0 \left[ (Y_0 - Y_1)k_x \sin(\alpha) + \sqrt{3}X_1 k_y \cos(\alpha) \right] \quad \textbf{[S4.3]}$$

$$V_{12}(\vec{k}) = -\sqrt{2}eAa_0 \begin{bmatrix} \tau\left[(Y_0^A - Y_1^A)k_x \sin(\alpha) + \sqrt{3}X_1^A k_y \cos(\alpha)\right] \\ +i\left[(X_0^B - X_1^B)k_x \cos(\alpha) + \sqrt{3}Y_1^B k_y \sin(\alpha)\right] \end{bmatrix} \quad \textbf{[S4.4]}$$

$$V_{21}(\vec{k}) = \left[V_{12}(\vec{k})\right]^* \quad \textbf{[S4.5]}$$

with

$$X(Y)_m = \langle d_{xy}(\vec{r}) | x(y) | d_{x^2-y^2}(\vec{r} - \vec{R_m}) \rangle, \ X(Y)_m^A = \langle d_{xy}(\vec{r}) | x(y) | d_{z^2}(\vec{r} - \vec{R_m}) \rangle,$$

$$X(Y)_m^B = \langle d_{x^2-y^2}(\vec{r}) | x(y) | d_{z^2}(\vec{r} - \vec{R_m}) \rangle, \ \vec{R_m} = a_0 \left( \cos\left(m\frac{2\pi}{3}\right)\hat{x} + \sin\left(m\frac{2\pi}{3}\right)\hat{y} \right).$$

and $[.]^*$ corresponds to complex conjugation.

If the wavefunction is tightly bound, it is reasonable to assume that $a$ is big enough such that, at the neighbor lattice site, the value of the radial part of the orbital function is constant: $d_\alpha(\vec{r} - \vec{R_m}) \approx CL_\alpha(\vec{r} - \vec{R_m})$ where $C$ is constant of space. Using this assumption, it is possible to show that, when the system has $C_{3v}$ symmetry, the following relationship is fulfilled:

$$(Y_0 - Y_1) \approx -\sqrt{3}X_1 \quad \textbf{[S4.6]}$$

$$(Y_0^A - Y_1^A) \approx \sqrt{3}X_1^A \quad \textbf{[S4.7]}$$

$$(X_0^B - X_1^B) \approx -\sqrt{3}Y_1^B \quad \textbf{[S4.8]}$$

B. Dispersion relation, Berry curvature, and Berry curvature dipole under perturbation

Instead of the perturbation Hamiltonian above, we consider the case of a general linear perturbation Hamiltonian:

$$P_L(\vec{k}) = \begin{bmatrix} a + a_x k_x' + a_y k_y' & (c + c_x k_x' + c_y k_y') - i(d + d_x k_x' + d_y k_y') \\ (c + c_x k_x' + c_y k_y') + i(d + d_x k_x' + d_y k_y') & b + b_x k_x' + b_y k_y' \end{bmatrix} \quad \textbf{[4.9]}$$

where $k_{x(y)}' = a_0 t k_{x(y)}$. The total Hamiltonian can be expressed as $H(\vec{k}) = H_0(\vec{k}) + P_L(\vec{k})$ where $H_0(\vec{k})$ is given by equation **[S4.1]** and $P_L(\vec{k})$ is described by equation **[S4.9]**. Solving the associated eigenvalue problem, we can obtain the dispersion relation of the conduction band and valence band and the associated Berry curvature as a function of $\vec{k}$.

One of the main interest here is the calculation of the Berry curvature and Berry curvature dipole. The conduction band Berry curvature, $\vec{\Omega_{CB}}(\vec{k})$ is defined as

$$\vec{\Omega_{CB}}(\vec{k}) = i \frac{\left\{ \langle \Phi_C | \frac{\partial H}{\partial k_x} | \Phi_V \rangle \langle \Phi_V | \frac{\partial H}{\partial k_y} | \Phi_C \rangle - \langle \Phi_C | \frac{\partial H}{\partial k_y} | \Phi_V \rangle \langle \Phi_V | \frac{\partial H}{\partial k_x} | \Phi_C \rangle \right\}}{E_g^2} \hat{z} \quad \textbf{[S4.10]}$$

where $|\Phi_{V(C)}\rangle$ is the $\vec{k}$-dependent quantum state of the valence(conduction) band electron and $E_g$ is the $\vec{k}$-dependent bandgap. The Berry curvature dipole, $\vec{\Lambda}^\Omega$ is defined as (5)

$$\vec{\Lambda}^\Omega \equiv \oint_{E_g(\vec{k})=E_{exc}} \vec{dk} \times \vec{\Omega_{CB}}(\vec{k}) \qquad \text{[S4.11]}$$

where the closed loop integral is performed along the path with $E_g(\vec{k})$ equal to particular interband transition $E_{exc}$. Hence, given the $E_g(\vec{k})$ derived from the dispersion relations and $\vec{\Omega_{CB}}(\vec{k})$, we can obtain the Berry curvature dipole. First, we plot the $\vec{\Omega_{CB}}(\vec{k})$ and $E_g(\vec{k})$ as a function of $\vec{k} \equiv (k_x, k_y)$ for a random linear perturbation in Fig. S7. As can be seen from this figure, both the Berry curvature and the bandgap have the same contour shape. This indicates that the Berry curvature dipole is zero.

We confirm this by solving the integral in equation **[S4.11]** analytically. In our analysis, we neglect the following terms in the expression $\vec{\Omega_{CB}}(\vec{k})$ of and $E_g(\vec{k})$: the terms that of second order or higher in the perturbation coefficients (denoted as $\{p\}$) and the terms that of third order or higher in $\vec{k}$. We obtain that the Berry curvature dipole is indeed zero.

Using this approximation, we can express the dispersion relations and the Berry curvature for the conduction band (the Berry curvature of the valence band is simply the negative of the one for conduction band). These quantities are given by

$$E_{CB}^\tau(k_x, k_y) = \left(\frac{\Delta}{2} + \frac{a_0^2 t^2 k^2}{\Delta}\right) + a + a_0 t(a_x k_x + a_y k_y) + o\left(\frac{\{p\}}{\Delta}\right)$$

$$\approx \left(\frac{\Delta}{2} + \frac{a_0^2 t^2 k^2}{\Delta}\right) + a + a_0 t(a_x k_x + a_y k_y) \qquad \text{[S4.12]}$$

$$E_{VB}^\tau(k_x, k_y) = -\left(\frac{\Delta}{2} + \frac{a_0^2 t^2 k^2}{\Delta}\right) + b + a_0 t(b_x k_x + b_y k_y) + o\left(\frac{\{p\}}{\Delta}\right)$$

$$\approx -\left(\frac{\Delta}{2} + \frac{a_0^2 t^2 k^2}{\Delta}\right) + b + a_0 t(b_x k_x + b_y k_y) \qquad \text{[S4.13]}$$

$$E_g^\tau(k_x, k_y) = E_{CB}^\tau(k_x, k_y) - E_{VB}^\tau(k_x, k_y)$$

$$\approx \Delta + \frac{2a_0^2 t^2 k^2}{\Delta} + (a-b) + a_0 t\left((a_x - b_x)k_x + (a_y - b_y)k_y\right) \qquad \text{[S4.14]}$$

$$\frac{\Omega_{CB}^{\tau}(\vec{k})}{a_0^2 t^2} = \tau\left(\left(-\frac{2}{\Delta^2}+\frac{12a_0^2 t^2 k^2}{\Delta^4}\right) - \frac{2(\tau c_x + d_y)}{\Delta^2} + \frac{4(a-b)}{\Delta^3} + \frac{6a_0 t\left((a_x - b_x)k_x + (a_y - b_y)k_y\right)}{\Delta^3}\right) + o\left(\frac{\{p\}}{\Delta^4}\right)$$

$$\Omega_{CB}^{\tau}(\vec{k}) \approx \frac{2\tau a_0^2 t^2}{\Delta^2}\left(\begin{array}{c}\left(-1+\frac{6a_0^2 t^2 k^2}{\Delta^2}\right) - (\tau c_x + d_y) + \frac{2(a-b)}{\Delta} \\ + \frac{3a_0 t\left((a_x - b_x)k_x + (a_y - b_y)k_y\right)}{\Delta}\end{array}\right) \quad \textbf{[S4.15]}$$

where $k^2 = k_x^2 + k_y^2$, index $CB(VB)$ means conduction (valence) band, $\tau$ is the valley index and it's equal to 1(-1) for K(K') valley, and $o(g)$ is Little-O notation, which means the terms are of smaller order compared to $g$. As can be seen from equation **[S4.14]** and **[S4.15]**, the value of $\vec{k}$ with the same bandgap $E_g(\vec{k})$ will also have the equal value of Berry curvature

Applying **[S4.2-S4.5]** to **[S4.12-S4.15]** we obtain

$$E_{CB}^{\tau}(k_x, k_y) = \frac{\Delta}{2} + \frac{a_0^2 t^2 k^2}{\Delta} \quad \textbf{[S4.16]}$$

$$E_{VB}^{\tau}(k_x, k_y) = -\frac{\Delta}{2} - \frac{a_0^2 t^2 k^2}{\Delta} + 2\tau eAa_0\left[(Y_0 - Y_1)k_x \sin(\alpha) + \sqrt{3}X_1 k_y \cos(\alpha)\right] \quad \textbf{[S4.17]}$$

$$E_g^{\tau}(k_x, k_y) = \Delta + \frac{2a_0^2 t^2 k^2}{\Delta} - 2\tau eAa_0\left[(Y_0 - Y_1)k_x \sin(\alpha) + \sqrt{3}X_1 k_y \cos(\alpha)\right] \quad \textbf{[S4.18]}$$

$$\Omega_{CB}^{\tau}(\vec{k}) \approx \frac{2\tau a_0^2 t^2}{\Delta^2}\left(\begin{array}{c}\left(-1+\frac{6a_0^2 t^2 k^2}{\Delta^2}\right) + \sqrt{2}eAa_0 \sin(\alpha)\left((Y_0^A - Y_1^A) + \sqrt{3}Y_1^B\right) \\ - \frac{6\tau eAa_0}{\Delta}\left[(Y_0 - Y_1)k_x \sin(\alpha) + \sqrt{3}X_1 k_y \cos(\alpha)\right]\end{array}\right) \quad \textbf{[S4.19]}$$

As can be seen from equation **[S4.16]** and **[S4.17]**, the conduction band energy remains unchanged while the valence band has an additional linear-in-momentum term that has an opposite sign for K and K' valley.

**V. CPGE current formulation due to asymmetric electron-hole relaxation**

In Section IV, we have shown that the Berry curvature dipole is zero. This indicates that the CPGE we observed does not fit into the framework of Berry curvature dipole-induced CPGE which was proposed in (5). However, we note that the close relationship between the Berry curvature dipole and CPGE is only valid when the relaxation time of the electron and hole are the same (5). This approximation is not acceptable for heterostructure since, due to the ultrafast interlayer charge transfer, the electron resides in MoS$_2$ while the hole is in WSe$_2$. The mobility and effective mass of electron in these two materials are different (6), which means that the carrier relaxation times are also different.

Here, we give a CPGE current density formulation considering the case where the relaxation time of the electron and hole are not the same. We start with the general expression of photocurrent (5, 7, 8)

$$\vec{J}_{ph} = -\frac{2\pi e^3}{\hbar m_e^2} \int \frac{d^2k}{(2\pi)^2} (f_V - f_C)(\tau_C \vec{v}_C - \tau_V \vec{v}_V) |\langle \Phi_C | \vec{A} \cdot \vec{p} | \Phi_V \rangle|^2 \delta(E_g - \hbar\omega)$$ **[S5.1]**

where $m_e$ is the bare electron mass, $f_{V(C)}$ is the electron occupation number in valence(conduction) band, $\tau_{V(C)}$ is the relaxation time of the valence(conduction) band, $\vec{v}_{V(C)} = \frac{1}{\hbar}\nabla_k E_{VB(CB)}$ is the electron velocity in valence(conduction) band, $\vec{A}$ is the vector potential of the optical excitation, $\vec{p} = \frac{m_e}{\hbar}\nabla_k H$ is the momentum operator, and $\hbar\omega$ is the excitation energy.

For circularly polarized excitation, $\vec{A} \cdot \vec{p} = \frac{E_\omega m_e}{i\hbar\omega\sqrt{2}}\left(\frac{\partial H}{\partial k_x} \pm i\frac{\partial H}{\partial k_x}\right)$ with $E_\omega$ is the electric field amplitude of the optical excitation and the $\pm$ depends on if it is left or right-handed circularly polarized excitation. For the case of small excitation intensity in undoped wide bandgap material, the valence band can be treated as fully occupied and the conduction band is empty. Hence, $f_V - f_C$ can be taken to be equal to 1. The CPGE can be obtained from $\vec{J}_{CPGE} = \vec{J}_{ph}^+ - \vec{J}_{ph}^-$ where the superscript indicates if it's left-handed (+) or right-handed (-) excitation. Using the description above and equation **[S4.10]** and **[S5.1]**, we obtain

$$\vec{J}_{CPGE} = \frac{e^3 E_\omega^2}{2\pi\hbar^3\omega^2}\int d^2k(\tau_C\vec{v}_C - \tau_V\vec{v}_V) E_g^2 \Omega_{CB}\delta(E_g - \hbar\omega)$$ **[S5.2]**

To analyze the effect of the relaxation time difference, we rewrite equation **[S5.2]** into

$$\vec{J}_{CPGE} = \frac{e^3 E_\omega^2}{2\pi\hbar^3\omega^2}\begin{pmatrix}\int d^2k\tau_V(\vec{v}_C - \vec{v}_V)E_g^2\Omega_{CB}\delta(E_g - \hbar\omega) \\ +\int d^2k(\tau_C - \tau_V)\vec{v}_C E_g^2\Omega_{CB}\delta(E_g - \hbar\omega)\end{pmatrix}$$ **[S5.3]**

The first term in **[S5.3]** is the one that depends on the Berry curvature dipole. This can be seen by rewriting it as

$$\vec{J}_{CPGE}^{(1)} = \frac{e^3 E_\omega^2 \tau_V}{2\pi\hbar^3\omega^2}\int d^2k(\vec{v}_C - \vec{v}_V)E_g^2\Omega_{CB}\delta(E_g - \hbar\omega)$$

$$= \frac{e^3 E_\omega^2 \tau_V}{2\pi\hbar^4\omega^2}\int d^2k\left(\frac{\partial E_g}{\partial k_x}\hat{x} + \frac{\partial E_g}{\partial k_y}\hat{y}\right)E_g^2\Omega_{CB}\delta(E_g - \hbar\omega)$$

$$= \frac{e^3 E_\omega^2 \tau_V}{2\pi\hbar^4\omega^2}\left(\int dk_y dE_g \Omega_{CB} E_g^2 \delta(E_g - \hbar\omega)\hat{x} + \int dk_y dE_g \Omega_{CB} E_g^2 \delta(E_g - \hbar\omega)\hat{y}\right)$$

$$= \frac{e^3 E_\omega^2 \tau_V}{2\pi\hbar^2}\left(\oint_{E_g=\hbar\omega} dk_y \Omega_{CB}\hat{x} + \oint_{E_g=\hbar\omega} dk_x \Omega_{CB}\hat{y}\right).$$ **[S5.4]**

Since the Berry curvature dipole is zero in our case, the CPGE depends only on the second term.

$$\vec{J}_{CPGE} = \frac{e^3 E_\omega^2(\tau_C - \tau_V)}{2\pi\hbar^3\omega^2}\int d^2k\vec{v}_C E_g^2\Omega_{CB}\delta(E_g - \hbar\omega)$$

$$= \frac{e^3 E_\omega^2 (\tau_C - \tau_V)}{2\pi \hbar^4 \omega^2} \int d^2k E_g^2 \Omega_{CB} \delta(E_g - \hbar\omega) \left( \frac{\partial E_{CB}}{\partial k_x} \hat{x} + \frac{\partial E_{CB}}{\partial k_y} \hat{y} \right)$$

$$= \frac{e^3 E_\omega^2 (\tau_C - \tau_V)}{2\pi \hbar^2} \oint_{E_g = \hbar\omega} dl \frac{\Omega_{CB} \left( \frac{\partial E_{CB}}{\partial k_x} \hat{x} + \frac{\partial E_{CB}}{\partial k_y} \hat{y} \right)}{\sqrt{\left( \frac{\partial E_g}{\partial k_x} \right)^2 + \left( \frac{\partial E_g}{\partial k_y} \right)^2}} \quad \text{[S5.5]}$$

Here, $dl = \sqrt{(dk_x)^2 + (dk_y)^2}$ and we have used the relation

$$\int_{\mathbf{R}^n} d\vec{r} f(\vec{r}) \delta(g(\vec{r})) = \int_{g^{-1}(0)} d\sigma(\vec{r}) \frac{f(\vec{r})}{|\nabla_r g|}$$

where the $\sigma(\vec{r})$ indicates a path where $g(\vec{r}) = 0$ holds.

Next, we calculate the $\overrightarrow{J_{CPGE}}$ for the case of a uniform in-plane electric field. To make the notation concise, we first use **[S4.12]** and **[S4.14-S4.15]** with $\{a, a_x, a_y, b\} = 0$ instead of **[S4.16]** and **[S4.18-S4.19]**. Appropriate substitution is then applied to $\{b_x, b_y, c_x, d_y\}$. In this case, the integration path can be written as

$$\left( k_x - \frac{\Delta b_x}{4a_0 t} \right)^2 + \left( k_y - \frac{\Delta b_y}{4a_0 t} \right)^2 = \left( \frac{\Delta}{2a_0 t} \sqrt{b_x^2 + b_y^2 + 2\left( \frac{\hbar\omega - \Delta}{\Delta} \right)} \right)^2 \quad \text{[S5.6]}$$

This is a circle centered at $\left( \frac{\Delta b_x}{4a_0 t}, \frac{\Delta b_y}{4a_0 t} \right)$ with radius $R = \frac{\Delta}{2a_0 t} \sqrt{b_x^2 + b_y^2 + 2\left( \frac{\hbar\omega - \Delta}{\Delta} \right)}$. After performing the integration in **[S5.5]**, we obtain

$$\overrightarrow{J_{CPGE}} = \frac{6\tau a_0 t e^3 E_\omega^2 (\tau_C - \tau_V)}{\hbar^3 \Delta} \left( \frac{\hbar\omega}{\Delta} - \frac{4 + \tau c_x + d_y}{3} \right) (b_x \hat{x} + b_y \hat{y}) \quad \text{[S5.7]}$$

Substituting the appropriate $\{b_x, b_y, c_x, d_y\}$ following **[S4.3, S4.4, S4.9]** and considering the contribution from both K and K' valley we obtain

$$\overrightarrow{J_{CPGE}} = \left( \frac{24 e^4 a_0 t}{\hbar^3 \Delta t} \left( \frac{\hbar\omega}{\Delta} - \frac{4t - \sqrt{2} eA \sin(\alpha) \left( (Y_0^A - Y_1^A) + \sqrt{3} Y_1^B \right)}{3t} \right) \right) E_\omega^2 A (\tau_C - \tau_V)$$

$$\left( (Y_0 - Y_1) \sin(\alpha) \hat{x} + \sqrt{3} X_1 \cos(\alpha) \hat{y} \right) \quad \text{[S5.8]}$$

which is of the same form as equation **[2]** in the main text.

To obtain the $I_{CPGE}$ we from the $\overrightarrow{J_{CPGE}}$, we use the approach explained in (9). In our case, this involve projecting the $\overrightarrow{J_{CPGE}}$ into a virtual electric field direction going from the (+) electrode to (-) electrode (i.e., the electric field direction when there is a positive bias). Doing this we obtain

$$I_{CPGE} = \beta\left((Y_0 - Y_1)\sin(\alpha)\hat{x} + \sqrt{3}X_1\cos(\alpha)\hat{y}\right) \cdot \left(\cos(\alpha)\hat{x} + \sin(\alpha)\hat{y}\right)$$

$$= \beta\frac{\left(Y_0 - Y_1 + \sqrt{3}X_1\right)}{2}\sin(2\alpha) \qquad \textbf{[S5.9]}$$

where $\beta = 1$ for (+) electrode and $\beta = -1$ for the (-) electrode. Equation **[S5.9]** is of the same form as equation **[3]** in the main text. From **[S5.8]**, we can see that the $\vec{J}_{CPGE}$ is linearly proportional to the applied electric field, excitation power, and the relaxation time asymmetry. It also gives the impression that pumping it with higher excitation energy will give bigger $\vec{J}_{CPGE}$. However, as discussed in Section I, for off-resonant excitation, the intervalley scattering is bigger. This results in smaller $\vec{J}_{CPGE}$.

Lastly, we note that, for $C_{3v}$ lattice symmetry, we can apply the relationship **[S4.6]** to equation **[S5.8-S5.9]**. This results in CPGE current density that is always orthogonal to the applied electric field, which implies zero CPGE current for this case. This is consistent with the symmetry of the lattice. In particular, due to the 3-fold rotational symmetry, the $I_{CPGE}$ for a system with $C_{3v}$ lattice symmetry must only have terms that can be expressed as $\sin(3n\alpha + \phi)$ with $n$ is an integer and $\phi$ is a phase constant. Hence, the $I_{CPGE}$ with the form as in **[S5.9]** must vanish for $C_{3v}$ lattice symmetry.

The nonzero CPGE current near the electrode edge can be attributed to the fact that, near the edge, the lattice symmetry is lower than $C_{3v}$. In the next section, we discuss the applicability of the model presented here for heterostructure case, including the possible symmetry breaking mechanisms.

## VI. Applicability of the model to the heterostructure case

In Section V, we have shown that a 2-band model with different hole and electron relaxation time can explain the observed CPGE phenomenon provided that the $C_{3v}$ lattice symmetry is broken at the excitation location. Here, we discuss the applicability of this model to the heterostructure case, including the possible symmetry breaking mechanism.

For the heterostructure case, there are two separate contributions for the conduction band electron current density: one from the WSe$_2$ conduction and one from the MoS$_2$ conduction band. In the 2-band model discussed before, the contribution from the MoS$_2$ conduction band is accounted for in the modification of the effective relaxation time of the conduction band electron. The main argument is that the relaxation time in MoS$_2$ and WSe$_2$ is different and, hence, the effective relaxation time of electron and hole cannot be treated to the same anymore.

However, not only that the relaxation time of the electron will be different, but also its velocity will also be different in MoS$_2$. There are two main reasons for this:
1. The lattice parameter $a_0$ and the hopping parameter integral $t$ are different in MoS$_2$ and WSe$_2$. The velocity is proportional to the parameter $(a_0 t)^2$. Using the value given in (3), we can calculate that, for the same $\vec{k}$ value, the electron velocity in MoS$_2$ is 80% smaller than the one in WSe$_2$.

2. Due to different lattice parameter and the nonzero twist angle, after the charge transfer, the electron will have different $\vec{k}$ value in MoS$_2$ conduction band (w.r.t to the K(K') valley in MoS$_2$) compared to the value before the charge transfer in WSe$_2$ conduction band (w.r.t to the K(K') valley in WSe$_2$).

We will show that, in the 2-band model, the first mechanism can be taken into account in the effective relaxation time calculation while the second mechanism does not have to be taken into account as long as the effect of trigonal warping can be neglected (parabolic band approximation still holds).

First, we note that the effect of the first mechanism is that the equation **[S5.3]** has to be rewritten as

$$\overrightarrow{J_{CPGE}} = \frac{e^3 E_\omega^2}{2\pi \hbar^3 \omega^2} \left( \begin{array}{l} \int d^2 k \tau_V \left( \overrightarrow{v_C^{eff}} - \overrightarrow{v_V} \right) E_g^2 \Omega_{CB} \delta\left(E_g - \hbar\omega\right) \\ + \int d^2 k \left( \tau_C - \tau_V \right) \overrightarrow{v_C^{eff}} E_g^2 \Omega_{CB} \delta\left(E_g - \hbar\omega\right) \end{array} \right) \quad \textbf{[S6.1]}$$

where $\overrightarrow{v_C^{eff}} \neq \overrightarrow{v_C} = \frac{1}{\hbar} \nabla_k E_{CB}$. However, the effect of the first mechanism is only to change the effective electron mass such that $\overrightarrow{v_C^{eff}} = \beta \overrightarrow{v_C}$. Hence, we can rewrite **[S6.1]** as

$$\overrightarrow{J_{CPGE}} = \frac{e^3 E_\omega^2}{2\pi \hbar^3 \omega^2} \left( \begin{array}{l} \int d^2 k \tau_V \left( \overrightarrow{v_C} - \overrightarrow{v_V} \right) E_g^2 \Omega_{CB} \delta\left(E_g - \hbar\omega\right) \\ + \int d^2 k \left( \beta\tau_C - \tau_V \right) \overrightarrow{v_C} E_g^2 \Omega_{CB} \delta\left(E_g - \hbar\omega\right) \end{array} \right) \quad \textbf{[S6.2]}$$

which is of the same form as **[S5.3]** if we do the substitution $\tau_C \leftarrow \beta\tau_C$. So the effect of the first mechanism can be included in the effective relaxation time calculation in the 2-band model.

Regarding the second mechanism, considering the C3v symmetry of the lattice, within the first Brillouin zone, we have to consider three different K$_M$-K$_W$ configuration (here subscript M and W mean MoS$_2$ and WSe$_2$ layer respectively). This is illustrated in Fig. S8. As shown in that figure, for a parabolic band, the lattice mismatch and the twist angle does not play any role. Hence, the effective conduction band is the same as the WSe$_2$ conduction band with different effective electron mass due to the influence of the MoS$_2$.

Next, we discuss here the possible mechanisms such that equation **[S4.6]** is invalid and, hence, the $\overrightarrow{J_{CPGE}}$ will have a nonzero component along the electric field. There are three potential contributions:
1. The residual unilateral strain due to the electrodes configuration breaks the C3v lattice symmetry.
2. The strain induced at the electrode-sample interface locally breaks the C3v lattice symmetry.
3. The nonuniform electric field (nonzero electric field gradient) contribution.

The first contribution has a quite obvious effect. Since the value of $\left(Y_0 - Y_1 + \sqrt{3}X_1\right) \neq 0$ will be the same all over the electrode-sample interface, it results in the $\sin(2\alpha)$ dependence observed in the experiment. The second and third contribution is more non-trivial. This is because, in these two cases, the value of $\left(Y_0 - Y_1 + \sqrt{3}X_1\right)$ will also depends on the value of $\alpha$. To deduce which symmetry breaking mechanism is more dominant, further experiments will be needed. This experiment can involve an electrode configuration similar to Corbino disk such that the value of $I_{CPGE}$ for various values of $\alpha$ can be obtained.

Lastly, we would like to address some assumptions used in the current discussion. Firstly, our discussion above only applies to the near-resonant excitation and linear perturbation (i.e., the electric field is small enough such that the nonlinear effect is negligible). When the excitation is off-resonant, the term that is of third order or higher in $\vec{k}$ has to be taken into account. These terms are responsible for the difference between the hole and electron effective mass and for the trigonal warping effect (4, 10). Including these terms may result in a nonzero Berry curvature dipole. However, since these terms are of a higher order, it is negligible compared to the effect discuss here. Similarly, the nonlinear perturbation effect is insignificant for a small electric field. The nonlinearity observed in Fig. 2*D* in the main text might be attributed to such higher order terms and nonlinear perturbation terms. Moreover, the effect of strain is limited to the $C_{3v}$ lattice symmetry breaking necessary to create nonzero $I_{CPGE}$. We argue that the residual unilateral strain should be quite small since we do not intentionally strain the sample. Hence, the band shift due to this strain is negligible. This is in contrast with the study of electric field effect in an intentionally strained sample (11). Finally, we emphasize here that the influence of charge transfer, lattice misalignment, and the band structure difference between MoS$_2$ and WSe$_2$ is included as effective electron relaxation time. The detail calculation of this effective relaxation time is beyond the scope of this report.

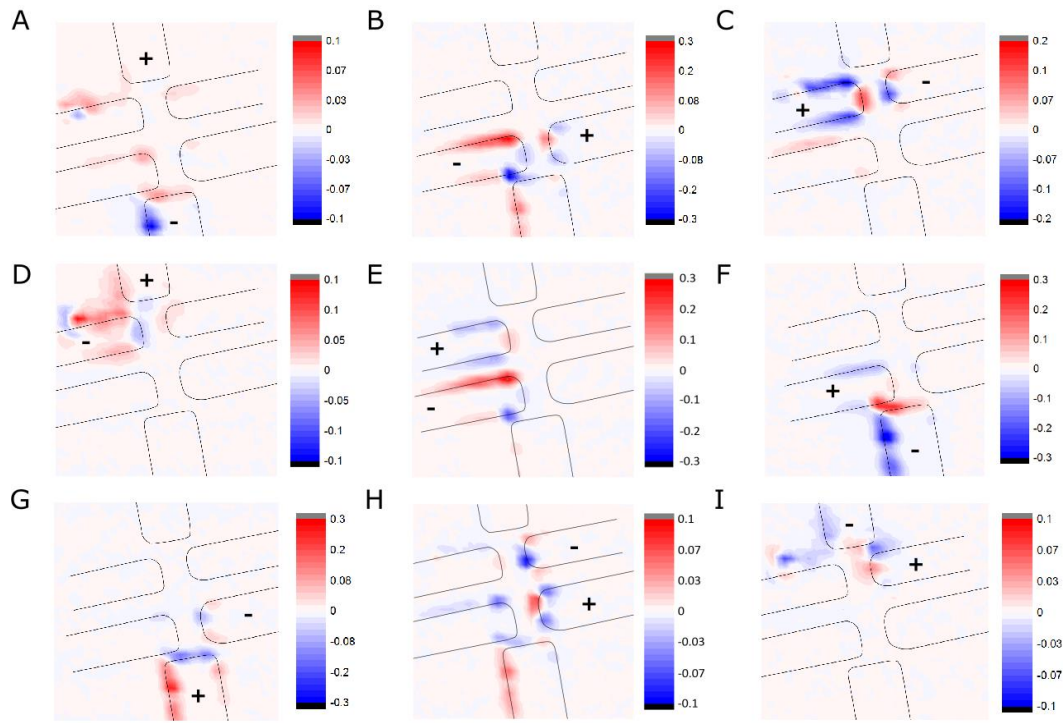

**Figure S1.** CPGE map at room temperature (295 K) for various electrode pair configuration. All of these data are obtained using 720 nm 115 μW excitation. The unit of the CPGE current is nA. In all of the cases, the CPGE is much more apparent when the excitation is near the electrode edge compared to other excitation locations.

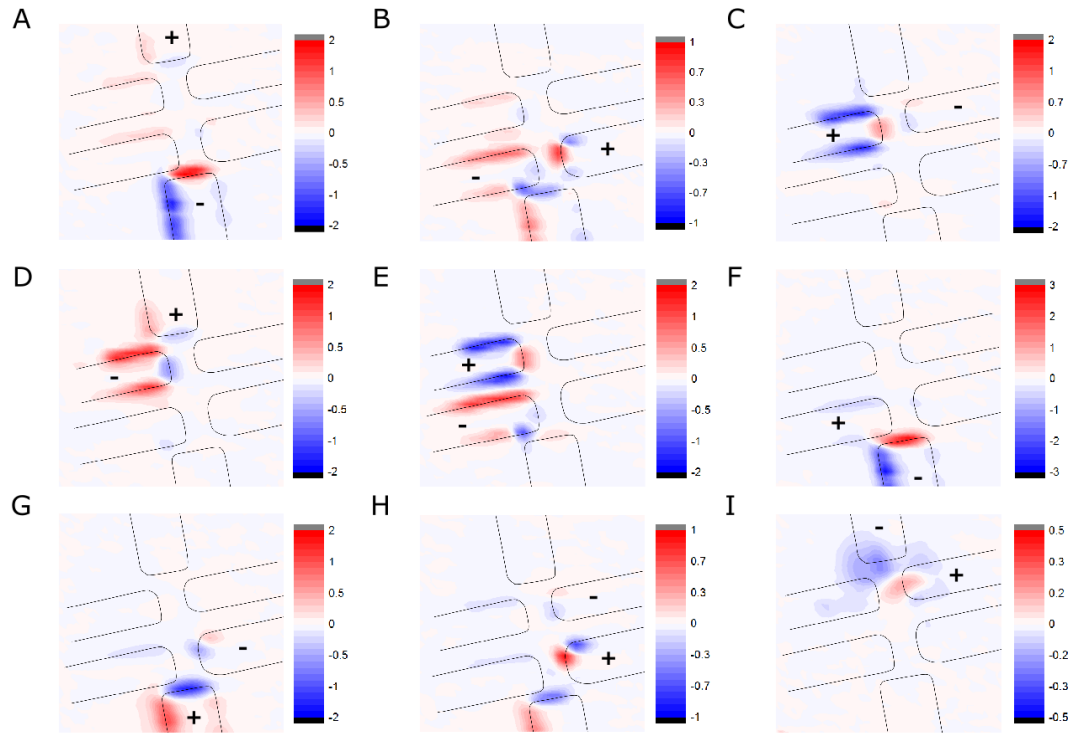

**Figure S2.** CPGE map at low temperature (140K) for various electrode pair configuration. All of these data are obtained using excitation with 720 nm wavelength and power between 90 - 100 μW. The unit of the CPGE current is nA. The CPGE map is qualitatively similar to the room temperature case but with much bigger CPGE.

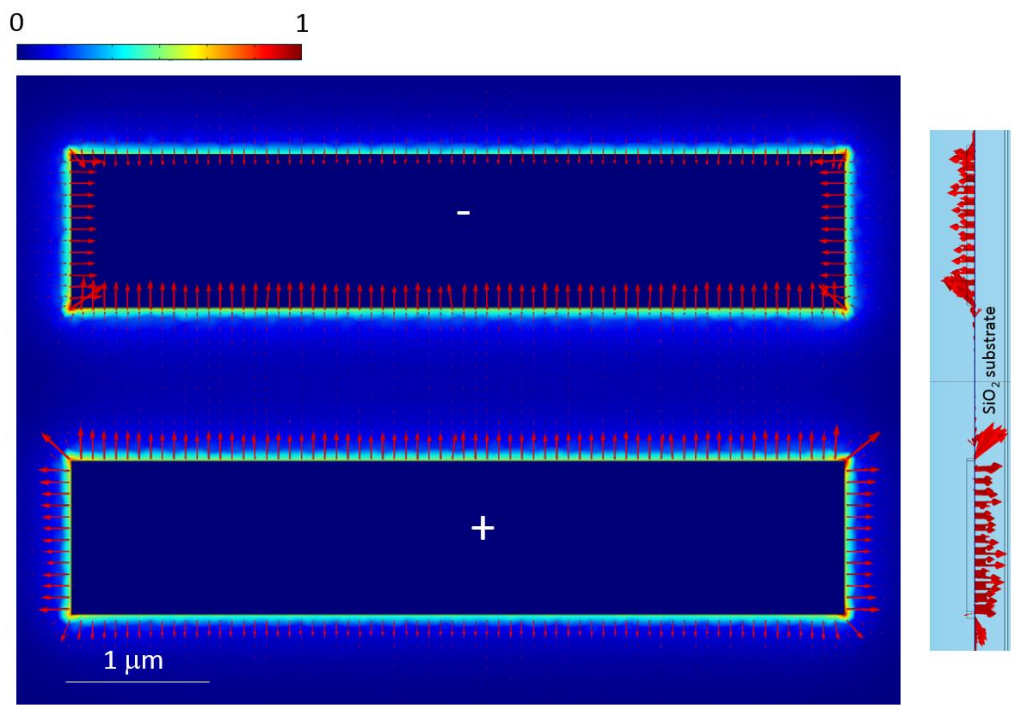

**Figure S3.** Illustration of the electric field due to the source-drain bias. Here the MoS$_2$/WSe$_2$ is treated as insulating as SiO2. Both the top view and side view of the sample are shown here. In the top view drawing, the color represents the normalized strength of the in-plane electric field while the arrows represent the in-plane electric field strength and direction. In the side view drawing, the arrows represent the total electric field strength and direction. The figure is obtained by using a COMSOL simulation. When the MoS$_2$/WSe$_2$ is not insulating, the electric field will be mainly in the in-plane direction.

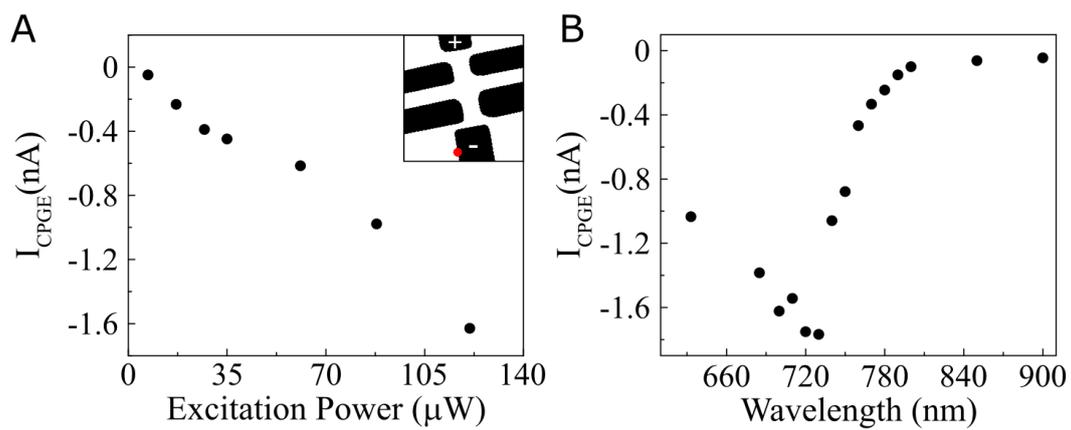

**Figure S4.** Power and wavelength dependence of CPGE at low temperature (140K). (*A*) CPGE as a function of excitation power. The excitation wavelength used here is 720 nm. The inset shows the excitation location and electrode configuration. (*B*) CPGE as a function of excitation wavelength. The excitation power is maintained around 140 – 155 µW. The excitation location and electrode configuration are the same as in Fig. S4(*A*).

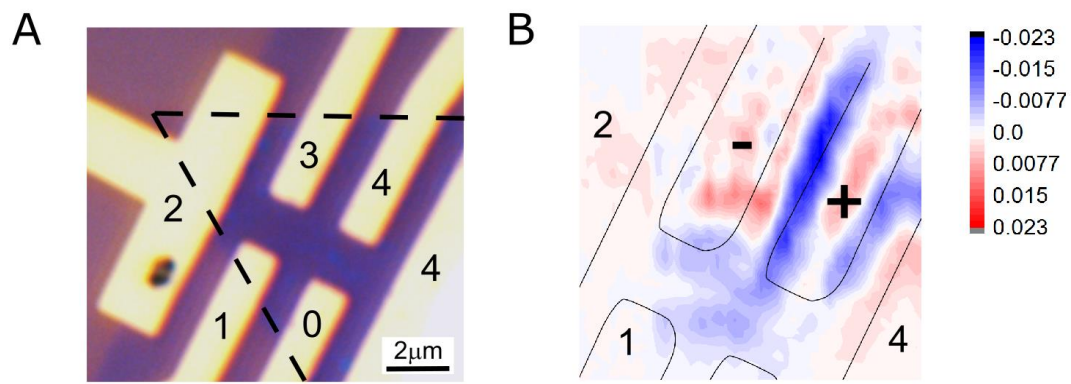

**Figure S5.** CPGE map at room temperature for another MoS$_2$/WSe$_2$ heterostructure device. (*A*) The layout of the device. The two electrodes numbered 4 are shorted to each other. The dashed line shows the boundary of the heterostructure. (*B*) CPGE map with source: electrode 3 and drain: electrode 4. There is no source-drain bias apply to the sample. The gate voltage of 5V is used. The unit of the CPGE current is nA.

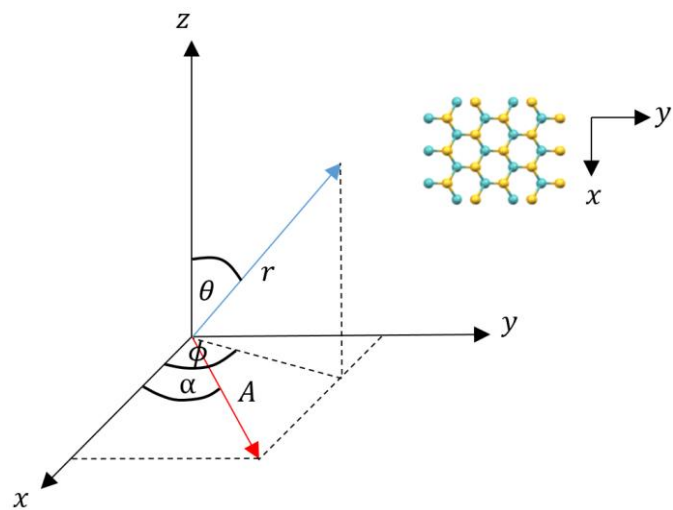

**Figure S6.** Space coordinate definition used in the report.

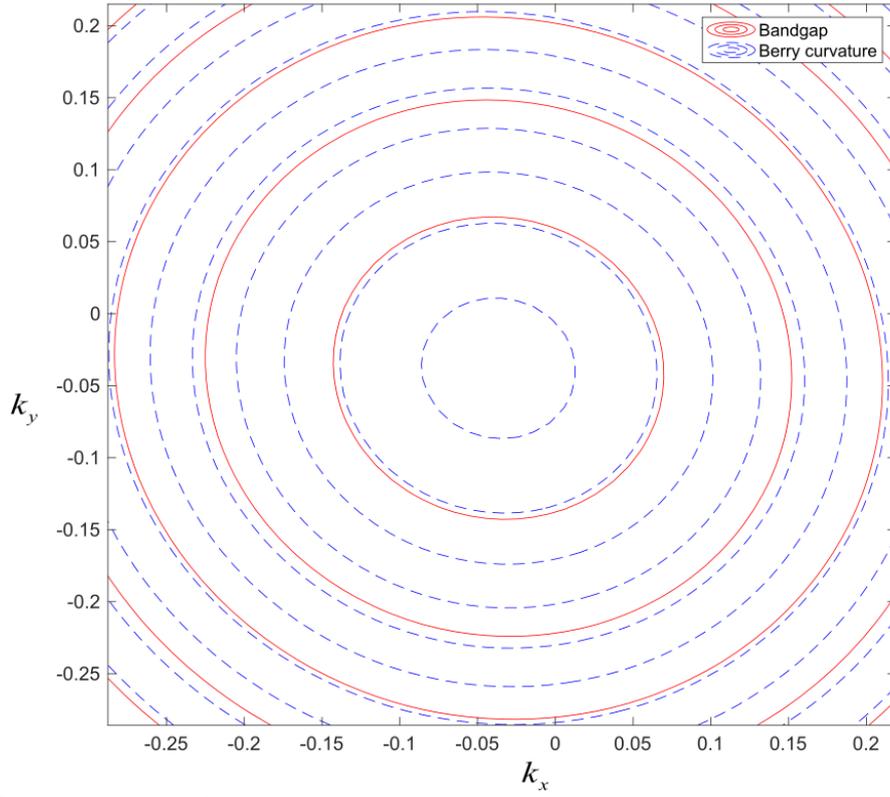

**Figure S7.** The contour of the bandgap and the berry curvature under small linear perturbation.

The values of the parameters used here are $\begin{Bmatrix} a, a_x, a_y, \\ b, b_x, b_y, \\ c, c_x, c_y, \\ d, d_x, d_y, \Delta \end{Bmatrix} = \begin{Bmatrix} 0.0569, 0.0469, 0.0012, \\ 0.0337, 0.0162, 0.0794, \\ 0.0311, 0.0529, 0.0166, \\ 0.0602, 0.0263, 0.0654, 1 \end{Bmatrix}$.

The unit is chosen such that $a_0 t = 1$ and $k_{x(y)} a_0 t$ have the same unit as $\Delta$.

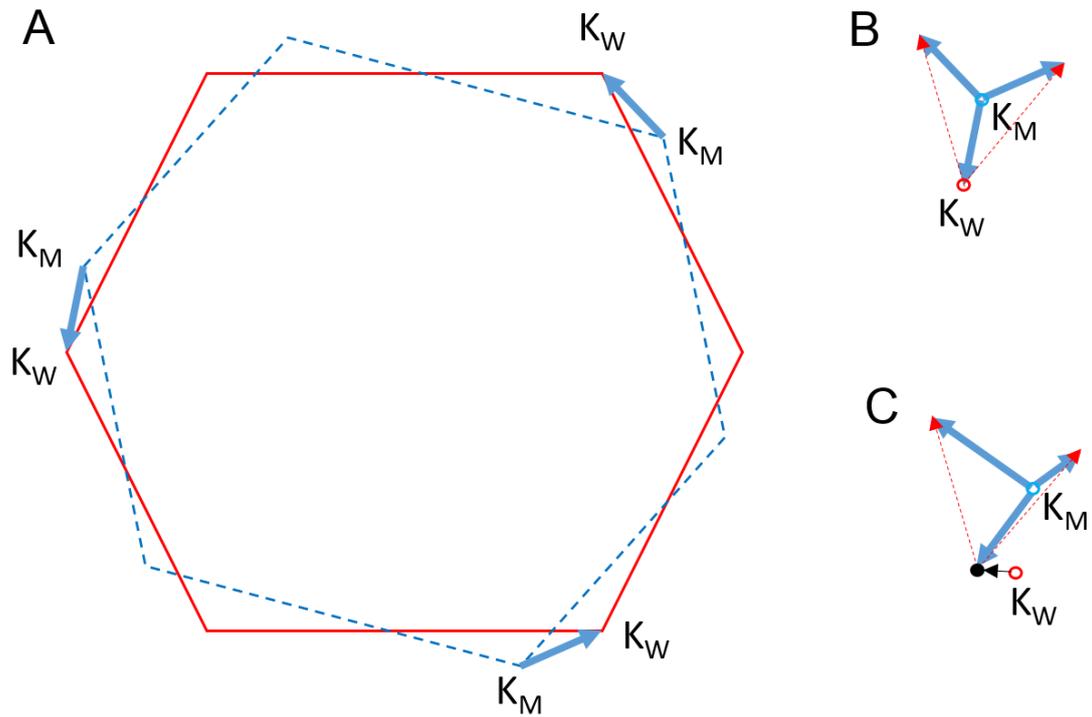

**Figure S8.** The effect of lattice mismatch and interlayer twist angle on the effective conduction band electron velocity. (A) The illustration of the first Brillouin zone of WSe$_2$ (red solid line hexagon) and that of MoS$_2$ (blue dotted line hexagon). The blue arrows represent the position of WSe$_2$ K valley with respect to MoS$_2$ K valley. For a parabolic band, the blue arrow is proportional to electron velocity. Adding all the blue arrows and divided by 3 will give the effective electron velocity. (B) For the case where the electron is originated from K valley before the charge transfer, the effective velocity is zero. (C) For the situation where the electron is not originated from K valley, the effective velocity has the same direction but with different magnitude. This shows that the effective conduction band is the same as the WSe$_2$ conduction band with different effective electron mass.

31